\documentclass[12pt,showpacs,preprintnumbers,superscriptaddress,amsmath,amssymb,nofootinbib]{revtex4}
\usepackage{amsmath, graphicx}
\usepackage{dcolumn}
\usepackage{bm}
\usepackage{color}
\usepackage{epstopdf}
\usepackage{eurosym}
\usepackage{multirow}
\usepackage{wrapfig}
\usepackage{amsfonts}
\usepackage{amsmath}
\usepackage{hyperref}
\usepackage{graphicx,color}
\usepackage{amssymb}
\usepackage[english]{babel}
\usepackage{graphicx}
\usepackage{epsfig}
\usepackage{bm}
\usepackage{longtable}
\usepackage{verbatim}
\usepackage{longtable}
\usepackage{color}
\usepackage{subfigure}
\usepackage{multirow}
\begin{document}
\newcommand{\hs}{\hspace*{0.5cm}}
\newcommand{\ms}{\hspace*{1cm}}
\newcommand{\vs}{\vspace*{0.5cm}}
\newcommand{\be}{\begin{equation}}
\newcommand{\ee}{\end{equation}}
\newcommand{\bea}{\begin{eqnarray}}
\newcommand{\eea}{\end{eqnarray}}
\newcommand{\ben}{\begin{enumerate}}
\newcommand{\een}{\end{enumerate}}
\newcommand{\bde}{\begin{widetext}}
\newcommand{\ede}{\end{widetext}}
\newcommand{\nn}{\nonumber}
\newcommand{\crn}{\nonumber \\}
\newcommand{\Tr}{\mathrm{Tr}}
\newcommand{\non}{\nonumber}
\newcommand{\noi}{\noindent}
\newcommand{\al}{\alpha}
\newcommand{\la}{\lambda}
\newcommand{\bet}{\beta}
\newcommand{\ga}{\gamma}
\newcommand{\va}{\varphi}
\newcommand{\om}{\omega}
\newcommand{\pa}{\partial}
\newcommand{\+}{\dagger}
\newcommand{\fr}{\frac}
\newcommand{\sq}{\sqrt}
\newcommand{\bc}{\begin{center}}
\newcommand{\ec}{\end{center}}
\newcommand{\Ga}{\Gamma}
\newcommand{\de}{\delta}
\newcommand{\De}{\Delta}
\newcommand{\ep}{\epsilon}
\newcommand{\varep}{\varepsilon}
\newcommand{\ka}{\kappa}
\newcommand{\La}{\Lambda}
\newcommand{\si}{\sigma}
\newcommand{\Si}{\Sigma}
\newcommand{\ta}{\tau}
\newcommand{\up}{\upsilon}
\newcommand{\Up}{\Upsilon}
\newcommand{\ze}{\zeta}
\newcommand{\ps}{\psi}
\newcommand{\Ps}{\Psi}
\newcommand{\ph}{\phi}
\newcommand{\vph}{\varphi}
\newcommand{\Ph}{\Phi}
\newcommand{\Om}{\Omega}

\title{Fermion Mass and Mixing in a Low-Scale Seesaw Model \\
based on the $S_4$ Flavor Symmetry}

\author{V. V. Vien }
\email{vovanvien@tdtu.edu.vn}
\affiliation{Theoretical Particle Physics and Cosmology Research Group, Advanced Institute of Materials Science,
Ton Duc Thang University, Ho Chi Minh City, Vietnam\\
and Faculty of Applied Sciences, Ton Duc Thang University, Ho Chi Minh City, Vietnam.}

\author{H. N. Long}
\email{hnlong@iop.vast.ac.vn}
\affiliation{Institute of Physics, VAST, 10 Dao Tan, Ba Dinh, Hanoi, Vietnam\\
and Bogoliubov Laboratory for Theoretical Physics,
Joint Institute for Nuclear Researches, 141980 Dubna, Moscow region, Russia.}

\author{A. E. C\'arcamo Hern\'andez}
\email{antonio.carcamo@usm.cl}
\affiliation{Universidad T\'{e}cnica Federico Santa Mar\'{\i}a and Centro Cient%
\'{\i}fico-Tecnol\'{o}gico de Valpara\'{\i}so, Casilla 110-V, Valpara\'{\i}so, Chile.}

\date{\today}

\begin{abstract}

We construct a low-scale seesaw model to generate the masses of active neutrinos based on $S_4$ flavor symmetry supplemented by the $Z_2 \times Z_3 \times Z_4 \times Z_{14}\times U(1)_L$ group, capable of reproducing the low energy Standard model (SM) fermion flavor data. The masses of the SM fermions and
 the fermionic mixings parameters are generated from a Froggatt-Nielsen mechanism after the spontaneous breaking of the
  $S_4\times Z_2 \times Z_3 \times Z_4 \times Z_{14}\times U(1)_L$ group.
The obtained values for the physical observables of the quark and lepton sectors are in good agreement with the most recent experimental data.
The leptonic Dirac CP violating phase $\de _{CP}$ is predicted to be $259.579^\circ$ and the predictions for the absolute neutrino masses in the model can also saturate the recent constraints.

\end{abstract}

\maketitle

\section{\label{intro}Introduction}

Despite its great success, the SM still has serious drawbacks such as the lack of mechanisms that explain the smallness of neutrino masses, the large hierarchy of charged fermion masses, the fermionic mixing angles, the leptonic CP violation, etc.
Another puzzle of the SM is that it does not explain why there are three generations of fermions. This puzzle can be addressed in the 3-3-1 models \cite{Antonio2017}. Hence, the neutrino masses and lepton mixings can be regarded as one of the most important evidence of physics beyond the SM. Among the possible extensions of the SM, discrete symmetries associated with the SM extensions are an useful tool to explain the observed pattern of SM fermion masses and mixing angles. According to the neutrino oscillation experimental data \cite{PDG2018}, the best fit values of neutrino mass squared differences and the leptonic mixing angles are
\bea
&&\sin^2(\theta_{12})=0.307 \pm 0.013,\,\,\, \sin^2(\theta_{13})=(2.18 \pm 0.07)\times 10^{-2},\crn
&& \sin^2(\theta_{23})= 0.536^{+0.023}_{-0.028} \hs \mathrm{ (Inverted \,\, order)},\crn
&& \sin^2(\theta_{23})= 0.512^{+0.019}_{-0.022} \hs  \mathrm{ (Normal \,\, order, octant \, I)},\label{PDG2018}\\
&& \sin^2(\theta_{23})= 0.542^{+0.019}_{-0.022} \hs  \mathrm{ (Normal \,\, order, octant \, II)},\crn
&& \De  m^2_{21}=(7.53\pm0.18)\times 10^{-5} \mathrm{eV}^2, \crn
&&\De  m^2_{32}= (-2.53\pm 0.05)\times 10^{-3}\mathrm{eV}^2 \,\, (S=1.2) \,\, \mathrm{ (Inverted \,\, order)},\crn
&& \De  m^2_{32}= (2.444\pm 0.034)\times 10^{-3}\mathrm{eV}^2 \hs\hs\hs\,\,\,\,  \mathrm{ (Normal \,\, order)}.\nn\eea
The large leptonic mixing angles given in Eq. (\ref{PDG2018}) are completely different from the quark mixing ones defined by the Cabibbo - Kobayashi - Maskawa (CKM) matrix \cite{CKM, CKM1} and this has stimulated works on flavor symmetries.

One of the most simplest possibilities to understand small non-zero neutrino masses is probably the seesaw  mechanism,
including type I, II, III and/or their combinations which  has been briefly reviewed in Ref. \cite{seesaw0}. However,  in these scenarios, the scale of the masses of the right-handed neutrinos should be very high that cannot be reached in the near future.
In the inverse-and linear seesaw mechanism \cite{seesaw1, seesaw2, seesaw3, seesaw4, seesaw5, seesaw6, seesaw7, seesaw8, seesaw9, seesaw10, seesaw11, seesaw13, seesaw14, seesaw15, seesaw16, seesaw17, seesaw18, seesaw19, seesaw20, seesaw21, seesaw22, seesaw23, seesaw24} the small neutrino masses arise as a result of new physics at $\mathrm{TeV}$ scale which may be probed at the LHC experiments. In such low-scale models, both renormalizable and non-renormalizable interactions are included, which can explain the fermion masses and mixings. In the basis ($\nu$, N, S), the neutrino mass matrix can be presented in the form of a $3\times 3$ block matrix where each element is a submatrix. Depending on the position of the zero elements in the mass matrix, active neutrinos can receive masses through inverse or/and linear seesaw mechanisms that all impose some elements of the mass matrix to be zero or very small and none
of them are forbidden by the SM symmetry, however, such terms can be avoided by introducing additional flavor symmetries.

In this paper we propose the possibility of predicting
fermion masses and mixing angles in the framework
of the low-scale seesaw mechanism with $S_4$ flavor symmetry. $S_4$ is the permutation group of four objects, which is also the
symmetry group of a cube. It has 24 elements divided into 5
conjugacy classes, with \underline{1}, \underline{1}$'$,
\underline{2}, \underline{3}, and \underline{3}$'$ as its 5
irreducible representations.
We will work in the basis in which $\underline{3},\underline{3}'$ are
real representations whereas $\underline{2}$ is complex. For the Clebsch-Gordan coefficients of $S_4$ group one can see, for instance, in the Ref. \cite{S4DLNV}.

The content of this paper goes as follows. In Sec. \ref{model} we present the necessary elements of the linear seesaw model under the $S_4$ symmetry as well as introduce the necessary
Higgs fields responsible for fermion masses and mixings. Section \ref{quark} deals with quark masses and mixings and
Section \ref{lepton} is devoted to lepton masses and mixings. We conclude in Section \ref{conclusion}.

\section{\label{model}The model}
We consider a three Higgs doublet model with several gauge singlet scalars, where the SM gauge symmetry is
 supplemented by the $S_4 \times Z_2 \times Z_3 \times Z_4 \times Z_{14}\times U(1)_L$ group. In this work, three left-handed leptons $\psi_L$ and three right-handed neutrinos $\nu_R$ as well as extra neural leptons $N_{L}, N_{R}, S_{L}, S_{R}$ are each put in one $S_4$ triplet while the first right-handed charged leptons $l_{1R}$ and the last two right-handed charged leptons $l_{2,3R}$ transform as $\underline{1}$ and $\underline{2}$ under $S_4$ symmetry, respectively.  For the quark sectors, all the families $q_{1L}, u_{1R}, d_{1R}$ are put in $\underline{1}'$ and $q_{2L}, q_{3L}, u_{2R}, u_{3R}, d_{2R}, d_{3R}$ transform as $\underline{1}$ under $S_4$.
The particle spectrum of our model and their assignments under the $SU(2)_L\times U(1)_L\times S_4 \times Z_2 \times Z_3 \times
Z_4 \times Z_{14}$ group is summarized in Tables \ref{quarkcont} and \ref{lepcont} where the numbered subscripts on fields in order define components of their
$S_4$ multiplet representations as well as the quantum numbers corresponding to other groups of the model. We use the $S_4$ discrete group
since it is the smallest non Abelian discrete group having irreducible triplet and doublet representations.
The discrete group $S_4$ is crucial to get a predictive fermion sector consistent with the low energy fermion flavor data. Extra symmetries $Z_2, Z_3, Z_4$ and $Z_{14}$ are additional introduced in order to get the desired structure of the fermion mass matrices that will be discussed in detail in Sec.\ref{lepton}.
\section{\label{quark}Quark masses and mixings}
The quarks content and the corresponding
scalar fields of the model, under the
$[ SU(2)_L, U(1)_L, S_4, Z_2, Z_3, Z_4, Z_{14}]$, is given in Table. \ref{quarkcont}.
\begin{table}[ht]
\centerline{\begin{tabular}{|c|c|c|c|c|c|c|c|c|c||c|c|c|c|c|c|c|c|}
\hline
 & $q_{1L}$ & $q_{2L}$ & $q_{3L}$ & $u_{1R}$ & $u_{2R}$ & $u_{3R}$ & $d_{1R}$ & $d_{2R}$ & $d_{3R}$ & $H$ & $H'$ & $H''$  &$\chi$
\\ \hline
$SU(2)_L$&2&2&2&1&1&1&1&1&1&2&2&2&1\\ \hline
$U(1)_L$&0&0&0&0&0&0&0&0&0&0&0&0&0\\ \hline
$S_4$ & $\underline{1}'$ & $\underline{1}$ & $\underline{1}$ & $\underline{1}'$ & $\underline{1}$ & $\underline{1}$ & $\underline{1}'$ & $\underline{1}$ & $\underline{1}$ & $\underline{1}$ & $\underline{1}'$& $\underline{1}$ & $\underline{1}$\\ \hline
$Z_2$ & 1& 1& 1& $-1$ & $-1$ & $-1$ & 1& 1& 1& -1 & -1 & 1& 1  \\ \hline
$Z_3$ & 1& 1& 1& 1 & 1 & 1 & 1& 1& 1& 1 & 1 & 1& 1  \\
 \hline
$Z_4$ & 1& 1& 1& 1& 1& 1& 1& 1& 1& $1$&$1$&$1$  & 1 \\ \hline
$Z_{14}$ & $e^{-\fr{3i\pi }{7}}$ & $e^{-\fr{2i\pi }{7}}$ & 1 & $e^{\fr{3i\pi }{7}}$ & $e^{\fr{2i\pi }{7}}$ & 1 & $e^{\fr{5i\pi }{7}}$ & $e^{\fr{3i\pi }{7}}$ & $e^{\fr{3i\pi }{7}}$ & 1 & 1 & 1  &$e^{-\fr{i\pi }{7}}$ \\ \hline
\end{tabular}}
\caption{\label{quarkcont} $SU(2)_L\times U(1)_L \times S_4 \times Z_2 \times Z_3 \times Z_4 \times Z_{14}$ assignments for quarks and scalars. }
\end{table}
The quark Yukawa terms invariant under the symmetries of the model under consideration take the form:
\bea
\mathcal{L}_Y^{(q) } &=&y_{11}^{(u) }\overline{q}%
_{1L}\widetilde{H}u_{1R}\fr{\chi ^{6}}{\La ^6}+y_{12}^{(
u) }\overline{q}_{1L}\widetilde{H'}u_{2R}\fr{\chi ^5 }{\La ^5 }%
+y_{13}^{(u) }\overline{q}_{1L}\widetilde{H'}u_{3R}\fr{\chi ^3 %
}{\La ^3 }  \crn
&&+y_{21}^{(u) }\overline{q}_{2L}\widetilde{H'}u_{1R}\fr{\chi
^5 }{\La ^5 }+y_{22}^{(u) }\overline{q}_{2L}\widetilde{H}%
u_{2R}\fr{\chi ^4 }{\La ^4 }+y_{23}^{(u) }\overline{q}%
_{2L}\widetilde{H}u_{3R}\fr{\chi ^2 }{\La ^2 } \crn
&&+y_{31}^{(u) }\overline{q}_{3L}\widetilde{H'}u_{1R}\fr{\chi
^3 }{\La ^3 }+y_{32}^{(u) }\overline{q}_{3L}\widetilde{H}%
u_{2R}\fr{\chi ^2 }{\La ^2 }+y_{33}^{(u) }\overline{q}%
_{3L}\widetilde{H}u_{3R} \crn
&&+y_{11}^{(d) }\overline{q}_{1L}H''d_{1R}\fr{\chi ^7}{\La
^{7}}+y_{22}^{(d) }\overline{q}_{2L}H''d_{2R}\fr{\chi ^5 }{%
\La ^5 }+y_{23}^{(d) }\overline{q}_{2L}H''d_{3R}\fr{\chi
^5 }{\La ^5 }  \crn
&&+y_{32}^{(d) }\overline{q}_{3L}H''d_{2R}\fr{\chi ^3 }{\La
^3 }+y_{33}^{(d) }\overline{q}_{3L}H''d_{3R}\fr{\chi ^3 }{%
\La ^3 }+H.c.
\label{Ly}
\eea

Note that the lightest of the physical neutral scalars states of $H$, $H'$, $H''$ is the SM-like $125$ GeV Higgs boson
discovered at the LHC. As indicated by Eq. (\ref{Ly}), the top quark mass mainly arises from the renormalizable quark Yukawa term involving $H$.
Thus the SM-like $125$ GeV Higgs predominantly arises from the CP even neutral  part of $H$. Furthermore, in view of the large amount
of free and uncorrelated parameters of the low energy scalar potential of the  model, there is a lot of freedom to adjust the required pattern of scalar masses,
thus allowing to safely assume that the remaining scalars are heavy and outside the LHC reach. In addition, the loop effects of the heavy scalars
 contributing to precision observables can be suppressed by making an appropriate choice of the free parameters in the scalar potential.
 These adjustments do not affect the physical observables in the quark and lepton sectors, which are determined mainly by the Yukawa couplings.

Assuming that the $SU(2)$ Higgs doublets $H$, $H'$, $H''$ do acquire vacuum expectation values (VEVs) at the electroweak symmetry breaking
 scale $v=246$ GeV and
  the gauge singlet scalar $\chi$ gets a VEV of the order of $\la\La$, with $\la=0.225$ - one of the Wolfenstein parameters and $\La$ - the model cutoff,
   we find that the SM quark mass matrices are given by:
\be
M_U =\left(
\begin{array}{ccc}
a_{11}^{(u) }\la^6 & a_{12}^{(u) }\la ^5
& a_{13}^{(u) }\la^3  \\
a_{12}^{(u) }\la^5  & a_{22}^{(u) }\la ^4
& a_{23}^{(u) }\la^2  \\
a_{13}^{(u) }\la ^3  & a_{23}^{(u) }\la ^2
& a_{33}^{(u) }%
\end{array}%
\right) \fr{v}{\sqrt{2}}, \ms M_D =\left(
\begin{array}{ccc}
a_{11}^{(d) }\la ^7  & 0 & 0 \\
0 & a_{22}^{(d) }\la ^5  & a_{23}^{(d) }\la
^5  \\
0 & a_{32}^{(d) }\la ^3  & a_{33}^{(d) }\la
^3 %
\end{array}%
\right) \fr{v}{\sqrt{2}} ,  \nn
\ee%
where
\bea
&&a_{11}^{(u)}\simeq 1.89391 + 0.404032i,\ms a_{12}^{(u)}=a_{21}^{(u)}
\simeq -1.42926 - 0.00898659i,\crn
&&a_{13}^{(u) }=a_{31}^{(u) }\simeq 0.704581 + 0.284696i,\ms a_{22}^{(u) }
\simeq 1.34823 - 0.00203271i ,   \crn
&&a_{23}^{(u)}=a_{32}^{(u)}\simeq -0.0703718 + 0.0148338i\, ,\,\,\,\,\, a_{33}^{(u) }\simeq 0.989285 - 0.000056837i,\crn
&&a_{11}^{(d)}\simeq 0.564554,\crn
&&a_{22}^{(d)}\simeq -0.534463,\ms a_{23}^{(d)}
=a_{32}^{(d)}\simeq 1.08071,\ms a_{33}^{(d)}\simeq 1.42119,
\eea
are $\mathcal{O}(1)$ dimensionless couplings. The values of the $\mathcal{O}(1)$ dimensionless couplings given above allows to successfully reproduce the experimental values of the quark mass spectrum, CKM parameters and Jarlskog invariant. As indicated by Table \ref{Observables}, our model is consistent with the low energy quark flavor data. Note that we use the $M_Z$-scale experimental values of the quark masses given by Ref. \cite{Bora:2012tx} (which are similar to those in \cite{Xing:2007fb}). The experimental values of the CKM parameters are taken from Ref. \cite{Olive:2016xmw}.
\begin{table}[tbh]
\begin{center}
\begin{tabular}{c|l|l}
\hline
Observable & Model value & Experimental value \\ \hline
$m_u (MeV)$ & \quad $1.11$ & \quad $1.45_{-0.45}^{+0.56}$ \\ \hline
$m_c (MeV)$ & \quad $639$ & \quad $635\pm 86$ \\ \hline
$m_t (GeV)$ & \quad $172.3$ & \quad $172.1\pm 0.6\pm 0.9$ \\ \hline
$m_d (MeV)$ & \quad $2.9$ & \quad $2.9_{-0.4}^{+0.5}$ \\ \hline
$m_s (MeV)$ & \quad $57.7$ & \quad $57.7_{-15.7}^{+16.8}$ \\ \hline
$m_b (GeV)$ & \quad $2.82$ & \quad $2.82_{-0.04}^{+0.09}$ \\ \hline
$\sin \theta^{(q)}_{12}$ & \quad $0.225$ & \quad $0.225$ \\ \hline
$\sin \theta^{(q)}_{23}$ & \quad $0.0421$ & \quad $0.0421$ \\ \hline
$\sin \theta^{(q)}_{13}$ & \quad $0.00365$ & \quad $0.00365$ \\ \hline
$J$ & \quad $3.18\times 10^{-5}$ & \quad $\left(3.18\pm 0.15\right)\times 10^{-5}$
\\ \hline
\end{tabular}%
\end{center}
\caption{Model and experimental values of the quark masses and CKM
parameters.}
\label{Observables}
\end{table}
\begin{figure}[tbp]
\includegraphics[width=1.05 \textwidth]{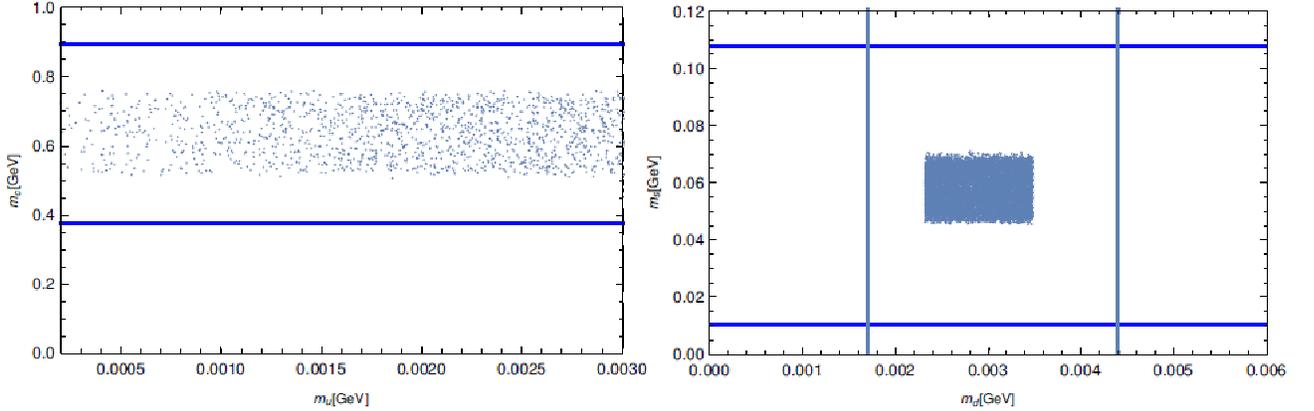}\newline
\caption{Correlations between the first and second generation SM quark
masses. The horizontal and vertical lines are the minimum and maximum values
of the second and first generation quark masses, respectively, inside the $3%
\protect\si$ experimentally allowed range.}
\label{c1}
\end{figure}
\begin{figure}[tbp]
\includegraphics[width=1.00\textwidth]{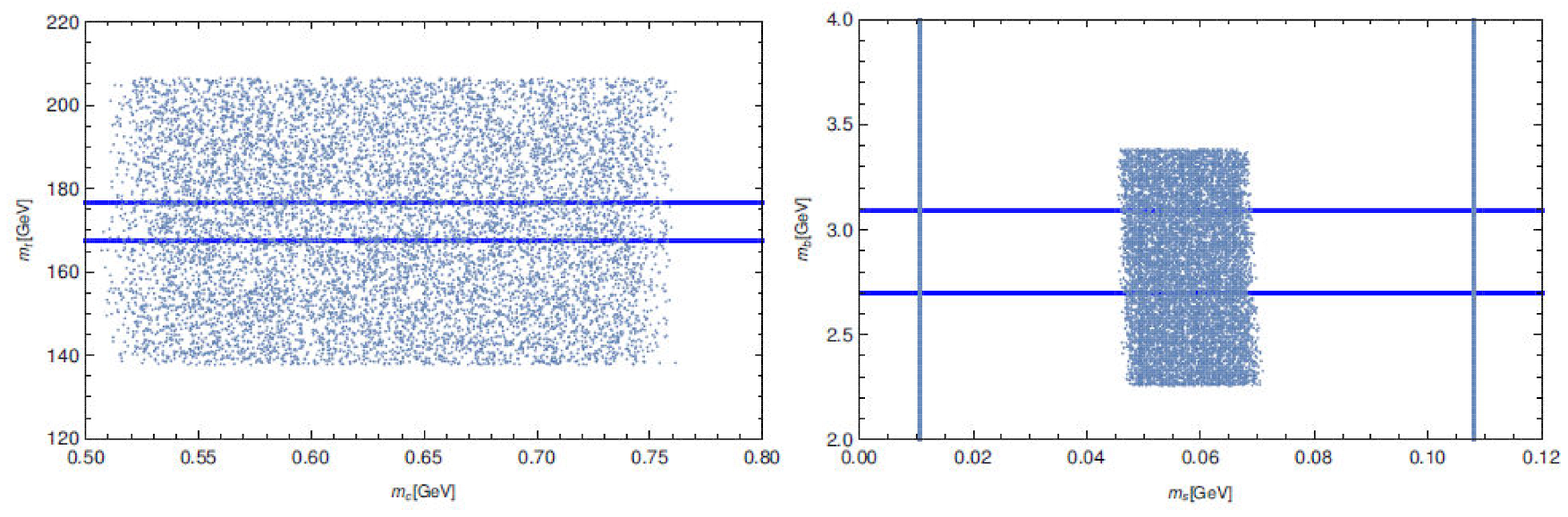}
\caption{Correlations between the second and third generation SM quark
masses. The horizontal and vertical lines are the minimum and maximum values
of the third and second generation quark masses, respectively, inside the $3%
\protect\si$ experimentally allowed range.}
\label{c2}
\end{figure}

With the aim to study the sensitivity of the obtained values for the SM quark
masses under variations around the best-fit values (maximum variation
around the $20\%$ of their best fit values), we show in Figs.~\ref{c1} and \ref{c2}, the correlations between the first and second as well as between third and second generation SM quark masses. We have found that such variations yield values for the SM quark masses inside the $3\si$ experimentally allowed range, with the exception of the top and bottom quark masses where the majority of points are outside the $3\si$ range. Consequently the quark sector model parameters feature some moderate amount of fine tuning. We have numerically checked that the up and down type quark sector parameters have to be varied in range around the $3\%$ and $4\%$ of their best fit values, respectively, in order to obtain all SM quark masses inside their $3\si$ experimentally allowed range.
\section{\label{lepton}Lepton masses and mixings}
\begin{figure}[t]
\vspace{0.5cm}
	\centering
	\includegraphics[width=10cm]{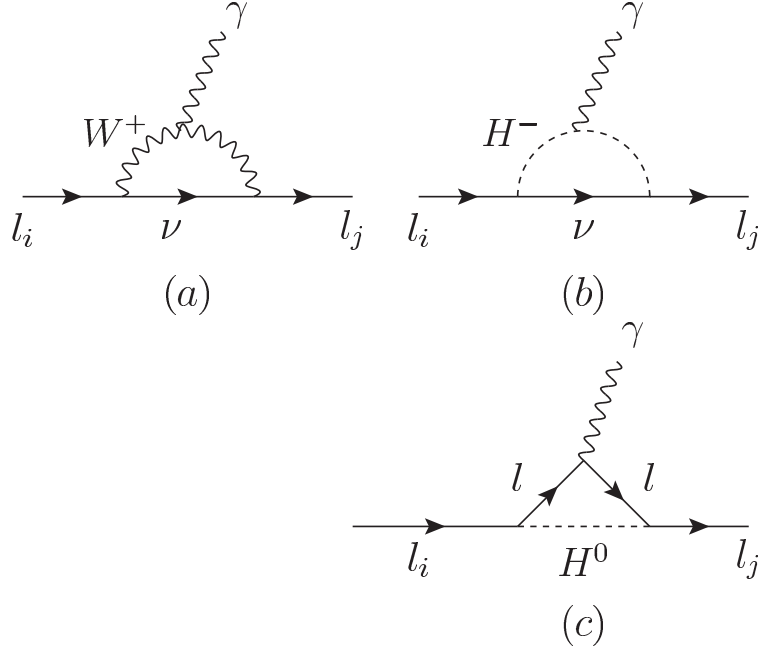}\\
	\caption{Feynman diagrams contributing to lepton  flavor changing decay. Here $i\neq j$ and $i = \tau, \mu $ and $ j= \mu \, , e $.}\label{decays4}
\end{figure}

The lepton fields and the corresponding scalars in lepton sectors, under the
$[ SU(2)_L, U(1)_L, S_4, Z_2, Z_3, Z_4 ]$, is given in Table \ref{lepcont}.
\begin{table}[ht]
\centerline{\begin{tabular}{|c|c|c|c|c|c|c|c|c||c|c|c|c|}
\hline
& $\psi_{L}$ & $l_{1R}$ & $l_{2,3R}$  & $\nu_{R}$& $N_L$ &$N_{R}$ & $S_L$ & $S_R$ & $\phi$ &$\vph $& $\xi$&$\rho$ \\ \hline
 $SU(2)_L$ & 2 & 1 &1&1 &1 &1 &1 &1 & 1 & 1 &1& 1  \\ \hline
$U(1)_L$ & 1 & 1 &1 &1 &1 &1 &1 &1 & 0 & 0 &0& 0  \\ \hline
$S_4$ & $\underline{3}$ & $\underline{1}$& $\underline{2}$  & $\underline{3}$ &$\underline{3} $ &$\underline{3}$& $\underline{3}$ & $\underline{3}$ &$\underline{3}$ &$\underline{3}$&$\underline{1}$&$\underline{1}$ \\ \hline
$Z_2$ & 1 & 1 & 1 & 1&1 & -1 & 1 &-1&  -1&-1 &-1 & -1  \\\hline
$Z_3$ & 1 & 1 & 1 & $\om $& $\om ^2$ & 1 & $\om ^2$ &1&  1&$\om  $&$\om $ & 1  \\  \hline
$Z_4$ & $i$&$i$&$i$ & 1& -$i$  & $i$& -$i$ & $i$ & 1 &-1&-1& $i$\\ \hline
\end{tabular}}
\caption{\label{lepcont}$SU(2)_L\times U(1)_L\times S_4 \times Z_2 \times Z_3 \times Z_4$ assignments for leptons and scalars. }
\end{table}
The lepton Yukawa terms invariant under the symmetries of the model are:
\bea
-\mathcal{L}_l&=&\fr{h_1}{\La }(\bar{\psi}_L\phi)_{\underline{1}} H l_{1R} +\fr{h_2}{\La }(\bar{\psi}_L \phi)_{\underline{2}}(H l_{R})_{\underline{2}} +
 \fr{h_3}{\La }(\bar{\psi}_L \phi)_{\underline{2}}(H' l_{R})_{\underline{2}}\crn
  &+&x_1 (\bar{\psi}_L N_R)_1  \widetilde{H}+x_2 (\bar{\psi}_L S_R)_1 \widetilde{H}+\fr{y_1}{\La} (\overline{N}_L \nu_R)_1  \xi \rho+\fr{y_2}{\La} (\overline{S}_L \nu_R)_1  \xi \rho \crn
 &+& z_1 (\overline{S}_L N_R)_1 \xi^\+  +z_2 (\overline{S}_L N_R)_{3_s}\vph ^\+  + t_1 (\overline{N}_L S_R)_1 \xi^\+ +t_2(\overline{N}_L S_R)_{3_s}\vph ^\+ \crn
 &+&w_1(\overline{N}_L N_R)_1 \xi^\+ +w_2(\overline{N}_L N_R)_{3_s}\vph ^\+ +w_3(\overline{S}_L S_R)_1 \xi^\+ +w_4(\overline{S}_L S_R)_{3_s}\vph ^\+
+H.c. \label{Ylep}\eea
In the case where $S_4 $ is spontaneously broken down to $\{\mathrm{identity}\}$ by the VEVs alignment
$\langle \phi_{1} \rangle =v,\, \langle \phi_{2} \rangle=v e^{i\al},\, \langle \phi_{3} \rangle =v e^{i\beta}$ and $\langle H \rangle= v_h$, $\langle H' \rangle= v'_h$ within the following expansions
 \be
 \phi_i = \langle \phi_i \rangle + \phi'_i \,\, (i=1,2,3),
 \label{ctl2}
 \ee
we get the lepton flavor  changing interactions as follows
  \bea
-\mathcal{L}^{clep}&\subset& \fr{h_1 v}{\La }(\bar{\nu}_{1L}H^+ + \bar{l}_{1L}H^0) l_{1R}+  \fr{h_2 v}{\La } (\bar{\nu}_{1L}H^+ + \bar{l}_{1L}H^0)l_{2R}\crn
&+&\fr{h_3 v}{\La }(\bar{\nu}_{1L}H'^+ +\bar{l}_{1L}H'^0)\bar{l}_{2R}+\fr{h_2 v}{\La }(\bar{\nu}_{1L}H^+ +\bar{l}_{1L}H^0)\bar{l}_{3R}\crn
&-&\fr{h_3 v}{\La }(\bar{\nu}_{1L}H'^+ + \bar{l}_{1L}H'^0)l_{3R}+\fr{h_1 v}{\La }(\bar{\nu}_{2L} H^+ + \bar{l}_{2L} H^0) l_{1R} + H.c \, .
 \label{ctl3}
 \eea
From \eqref{ctl3}, it follows that, in the model under consideration, the usual Yukawa couplings are associated with  the factor
$\fr v \La$ and the lepton flavor  changing decays consist of the contribution of  three Feynman diagrams as in Fig. \ref{decays4}.
\\
The current experimental data on lepton flavor  changing decays read \cite{Olive:2016xmw}: $\text{Br}(\mu^- \to e^-\ga) < 4.2\times 10^{-13}, \text{Br}(\tau^- \to e^-\ga) < 3.3\times 10^{-8}$ and $\text{Br}(\tau^- \to \mu^-\ga) < 4.4\times 10^{-8}$.
The partial decay width  is given by \cite{br1,br2}
\be
\Ga(l_i\rightarrow l_j\ga) = \fr{(m^2_i-m^2_j)^3}{16\pi m^3_i}\left(|C_L|^2+|C_R|^2\right), \label{ct281}\ee
where the above form factors  $C_L$ and $C_R$ are determined from the process amplitude \cite{br1,br2}
\bea
\mathcal{M} &=& 2(p_i\cdot\ep)
\left[C_L\bar{u}_j(p_j)P_Lu_i(p_i)+C_R\bar{u}_j(p_j)P_Ru_i(p_i)\right]\crn
 &-& (m_iC_R + m_jC_L) \bar{u}_j(p_j)/\!\!\!\ep P_L u_i(p_i) - (m_iC_L + m_jC_R)\bar{u}_j(p_j)/\!\!\!\ep P_R u_i(p_j),
\label{ct282}\eea
For the case $m_i \gg m_j$, we get
\bea
\text{Br}(l_i \to l_j \ga) = \fr{12\pi^2}{G_F^2}\left(|D_L|^2+|D_R|^2\right)\text{Br}(l_i \to l_j \bar{\nu}_j \nu_i),
\label{ct283}
\eea
where $G_F=g^2/(4\sqrt{2}m^2_W)$.
In the model under consideration, one has \cite{br2,br3}
\be D_L \propto  \fr{ v}{M_H\La }  \mathcal{O}(m_j/m_i)\, , \hs  D_R \propto  \fr{ v}{M_H\La }
\label{ct284}
\ee
where $M_H$ is the mass scale of the heavy scalars (which provide the dominant contributions to the LFV decays) running in the internal lines of the loop.
For further details on the  form factors $D_{L,R}$, the
reader is referred to Refs. \cite{br1,br2,br3,br4}. \\
Combining \eqref{ct283} and \eqref{ct284}, we see that the lepton flavor changing processes in this model are suppressed by the factor $\fr{v}{\La G_F^2M_H^2}$ associated with the above mentioned small Yukawa couplings and the large mass scale of the heavy scalars running in the internal lines of the loop.

Let us turn into lepton mass  issue. From \eqref{Ylep}, the   lepton mass terms read
\bea
-\mathcal{L}^{\mathrm{mass}}_{cl}
&=&\fr{ v_1}{\La }h_1v_h \bar{l}_{1L} l_{1R}+\fr{v_1}{\La }\left(h_2v_h+h_3v'_h\right)\bar{l}_{1L} l_{2R}+\fr{v_1}{\La }\left(h_2v_h -h_3v'_h\right)\bar{l}_{1L} l_{3R}\crn
 &+&\fr{ v_2}{\La }h_1v_h\bar{l}_{2L} l_{1R}+\fr{v_2}{\La }\left(h_2v_h+h_3v'_h\right)\om\bar{l}_{2L} l_{2R}+\fr{v_2}{\La }\left(h_2v_h-h_3v'_h\right)\om^2\bar{l}_{2L}l_{3R}\crn
 &+&\fr{v_3}{\La }h_1 v_h\bar{l}_{3L} l_{1R}+\fr{v_3}{\La }\left(h_2v_h+h_3v'_h\right)\om^2\bar{l}_{3L} l_{2R}+\fr{v_3}{\La }\left(h_2v_h-h_3v'_h\right)\om\bar{l}_{3L}l_{3R} +H.c\crn
&\equiv& (\bar{l}_{1L} \hs \bar{l}_{2L}\hs \bar{l}_{3L})
M_l (l_{1R}\hs l_{2R}\hs l_{3R})^T+H.c,
\eea
where the mass matrix for charged leptons is given by:
\bea M_l=
\fr v \La \left(%
\begin{array}{ccc}
  h_1v_h& h_2v_h+h_3v'_h & h_2v_h-h_3v'_h  \\
   h_1v_h e^{i\al} & (h_2v_h+h_3v'_h)e^{i\al} \om^2 & (h_2v_h-h_3v'_h)e^{i\al}\om \\
  h_1v_h e^{i\beta} & (h_2v_h+h_3v'_h)e^{i\beta} \om & (h_2v_h-h_3v'_h)e^{i\beta} \om^2\\
\end{array}%
\right).\label{Mltq}\eea
This matrix can be diagonalized as,
\bea U^\dagger_L M_l U_R=\fr{\sqrt{3} v}{\La } \mathrm{diag}( h_1 v_h, \, h_2v_h-h_3v'_h, \, h_2v_h+h_3v'_h )\equiv
 \mathrm{diag}(m_e, \, m_\mu , \,m_\tau),\label{Mld}\eea
where \bea U_L&=&\fr{1}{\sqrt{3}}\left(%
\begin{array}{ccc}
  1 &0 &0 \\
  0 &e^{i\al} &0 \\
  0 &0 &e^{i\beta} \\
\end{array}%
\right)\left(%
\begin{array}{ccc}
  1 &\,\,\, 1 &\,\,\, 1 \\
  1 &\,\,\, \om^2 &\,\,\, \om \\
  1 &\,\,\, \om &\,\,\, \om^2 \\
\end{array}%
\right),\hs U_R=1, \label{Uclep}\\
m_e &=&\fr{\sqrt{3} v}{\La } h_1 v_h,\hs  m_{\mu,\tau}=\fr{\sqrt{3} v}{\La } \left(h_2v_h \pm h_3v'_h\right) , \label{memt}\eea
where $\om=e^{i2\pi/3}$ is the cube root of unity.

The best fit values for the masses of charged-leptons are given in Ref. \cite{PDG2018}: $m_e \simeq 0.51099 \, \textrm{MeV}$, $\, m_\mu \simeq 105.65837 \, \textrm{MeV}$, $\,m_\tau \simeq 1776.86 \,\textrm{MeV}$. Then, we find the relations $\fr{h_3}{h_2} \simeq \fr{v_h}{v'_h},\, \fr{h_2}{h_1}\simeq 10^3$.

We also assume that in the neutrino sector, the $S_4$ discrete group is spontaneously broken down to the Klein four group $\mathcal{K}$ by the VEV alignment $\langle\vph  \rangle= (0, v_\vph , 0)$  of $\vph $ and the VEVs of $\xi, \rho$ as $\langle \xi \rangle= v_\xi ,\,  \langle\rho \rangle= v_\rho$. In this case, the neutrino mass matrices become
 \bea
 m_{\nu N} &=& x_1 v_h \textbf{I}\equiv a_1 \textbf{I}, \hs M_{\nu S} = x_2 v_h \textbf{I}\equiv a_2 \textbf{I}, \label{submatrix1}\\
m'_{\nu N}&=&  \fr{y_1 v_\xi  v_\rho}{\La} \textbf{I}\equiv b_1 \textbf{I}, \,\,  M'_{\nu S} =  \fr{y_2 v_\xi v_\rho}{\La} \textbf{I}\equiv b_2 \textbf{I}, \label{submatrix2}\\
 M'_{NS}&=& \left(%
\begin{array}{ccc}
  z_1 v_\xi & 0  & z_2 v_\vph  \\
  0 & z_1 v_\xi&0 \\
  z_2 v_\vph  & 0& z_1 v_\xi \\
\end{array}%
\right)\equiv  \left(%
\begin{array}{ccc}
  c_1 & 0 & c_2  \\
  0 & c_1 & 0 \\
  c_2 & 0  & c_1 \\
\end{array}%
\right), \label{submatrix3}\\
M_{NS} &=&\left(%
\begin{array}{ccc}
  t_1 v_\xi &0 & t_2 v_{\vph }  \\
  0 & t_1 v_\xi & 0 \\
  t_2v_{\vph } &0 & t_1 v_\xi \\
\end{array}%
\right)\equiv  \left(%
\begin{array}{ccc}
  d_1 & 0 & d_2  \\
  0 & d_1 & 0 \\
  d_2 & 0 & d_1 \\
\end{array}%
\right),\label{submatrix4}\\
M_{NN}&=&\left(%
\begin{array}{ccc}
  w_1 v_\xi &0 & w_2 v_{\vph }  \\
  0 & w_1 v_\xi & 0 \\
  w_2 v_\vph  &0 & w_1 v_\xi \\
\end{array}%
\right)\equiv  \left(%
\begin{array}{ccc}
  g_1 & 0 & g_2  \\
  0 & g_1 & 0 \\
  g_2 & 0 & g_1 \\
\end{array}%
\right), \label{submatrix5}\eea

\bea
 M_{SS}&=&\left(%
\begin{array}{ccc}
  w_3 v_\xi &0 & w_4 v_\vph   \\
  0 & w_3 v_\xi & 0 \\
  w_4 v_\vph  &0 & w_3 v_\xi \\
\end{array}%
\right)\equiv  \left(%
\begin{array}{ccc}
  g_3 & 0 & g_4  \\
  0 & g_3 & 0 \\
  g_4 & 0 & g_3 \\
\end{array}%
\right).\label{submatrix6}\eea
Let us note that the matrices given by Eqs. (\ref{submatrix1}) - (\ref{submatrix6}) are all symmetric and $m_{\nu N}$, $M_{\nu S}$, $ M'_{NS}$, $ M_{NS}$, $ M_{NN}$, $ M_{SS}$ are respectively generated from the renormalizable Yukawa interactions $x_1 (\bar{\psi}_L N_R)_1  \widetilde{H}$, $ x_2 (\bar{\psi}_L S_R)_1 \widetilde{H}$, $ \left\{z_1 (\overline{S}_L N_R)_1 \xi^\+, z_2 (\overline{S}_L N_R)_{3_s}\vph ^\+\right\}$, $ \left\{t_1 (\overline{N}_L S_R)_1 \xi^\+ , t_2(\overline{N}_L S_R)_{3_s}\vph ^\+ \right\}$, $\left\{w_1(\overline{N}_L N_R)_1 \xi^\+ , w_2(\overline{N}_L N_R)_{3_s}\vph ^\+\right\}$, $\left\{w_3(\overline{S}_L S_R)_1 \xi^\+, w_4(\overline{S}_L S_R)_{3_s}\vph ^\+\right\}$, whereas $m'_{\nu N}$ and $M'_{\nu S}$ arise from the non-renormalizable Yukawa interactions $\fr{y_1}{\La} (\overline{N}_L \nu_R)_1  \xi \rho$ and $\fr{y_2}{\La} (\overline{S}_L \nu_R)_1  \xi \rho$, respectively.

In this work, we introduce the $Z_2\times Z_3\times Z_4\times Z_{14}\times U(1)_L$ symmetry\footnote{All the lepton fields and the corresponding scalars in Table \ref{lepcont} carry the same charged (+ 1) under $Z_{14}$ which is not necessary to write out here.}, which in addition to the $S_4$ symmetry  to prevent some Yukawa interactions thus giving rise to the predictive textures for the neutrino sector shown in Eqs. (\ref{submatrix1}) - (\ref{submatrix6}). For instance, since the product of two $S_4$ triplets contains a $S_4$ triplet, the coupling $\overline{\psi}_L N_R$ can transform under $S_4\times Z_2\times Z_3\times Z_4\times Z_{14}\times U(1)_L$ as $\sim (\underline{3}\otimes \underline{3}, -1, 1,1,1,0)$, which implies that in order to generate the mass matrix $m_{\nu N}$, one needs  one $S_4$ singlet transforming as (\underline{1}, -1, 1,1,1,0), in order to build an invariant under all given symmetries. For the known scalars, $(\overline{\psi}_L N_R) \widetilde{H}'$ is forbidden by the $S_4$ symmetry, $(\overline{\psi}_L N_R) \widetilde{H}''$ is prevented by the $Z_2$ symmetry, $(\overline{\psi}_L N_R) \chi$ is not allowed by the $Z_2, Z_{14}$ and $SU(2)_L$ symmetries, whereas $(\overline{\psi}_L N_R) \xi$ is forbidden by $Z_3$ and $Z_4$ symmetries and $(\overline{\psi}_L N_R) \rho$ is prevented by the $Z_4$ symmetry. Consequently, there is only one term involving the fields $\psi_L$, $N_R$ and $H$, invariant under the $S_4\times Z_2\times Z_3\times Z_4\times Z_{14}\times U(1)_L$ symmetry, which corresponds to $x_1 (\bar{\psi}_L N_R)_1  \widetilde{H}$ as in Eq. (\ref{Ylep}) that provide a simple form of $m_{\nu N}$ as indicated by Eq. (\ref{submatrix1}). The situation is similar for the remaining couplings that generate the other mass matrices given in Eqs.(\ref{submatrix1}) - (\ref{submatrix6}).

In the basis ($\nu$ , N, S), the full neutrino mass matrix predicted by our model takes the form:
  \bea
 M_{\mathrm{eff}}&=&  \left(%
 \begin{array}{ccc}
  0& m_{\nu N} & M_{\nu S} \\
 m'_{\nu N} & M_{NN} &M_{NS} \\
 M'_{\nu S} &  M'_{NS}  &M_{SS} \\
\end{array}%
\right). \label{Meff0}\eea
The light active neutrino masses are obtained by diagonalizing the matrix given by Eq. (\ref{Meff0}) and this is done by introducing the following matrices
\bea
M_D&=&(m_{\nu N} \,\,\, M_{\nu S}),\,\, M^T_D= \left(%
 \begin{array}{c}
  m'_{\nu N}\\
 M'_{\nu S} \\
\end{array}%
\right), \,\, M_R= \left(%
 \begin{array}{cc}
M_{NN} & M_{NS} \\
M'_{NS}  & M_{SS} \\
\end{array}%
\right).\nn
   \eea
The effective neutrino mass matrix $M_{\mathrm{eff}}$ in Eq. (\ref{Meff0}) can be rewritten in the form:
\bea
 M_{\mathrm{eff}}&=& \left(%
 \begin{array}{ccc}
  0&M_D \\
 M^T_D & M_R \\
\end{array}%
\right), \label{Meff}\eea
which is similar to the one resulting from a type-I seesaw mechanism. Then, the light active neutrino mass matrix takes the form:
 \bea
    m_\nu&=&- M_D M^{-1}_R M^T_D
    =-m_{\nu N}M'^{-1}_{NS}M'_{\nu S}-M_{\nu S}M^{-1}_{NS}m'_{\nu N}\crn
 &-& M_{\nu S}M^{-1}_{SS}M'_{\nu S}+m_{\nu N}M^{-1}_{NN}M_{NS}M^{-1}_{SS}M'_{\nu S}+M_{\nu S}M^{-1}_{SS}M'_{NS}M^{-1}_{NN}m'_{\nu N} \crn
 &+&M_{\nu S}M^{-1}_{NS}M_{NN}M'^{-1}_{NS} M'_{\nu S}-m_{\nu N}M^{-1}_{NN} M_{NS}M^{-1}_{SS}M'_{NS}M^{-1}_{NN}m'_{\nu N}.  \label{mnu}
   \eea
   Replacing Eqs. (\ref{submatrix1}) - (\ref{submatrix6}) in Eq. (\ref{mnu}) yields the following mass matrix for light active neutrinos:
   \bea
   m_\nu &=& \left(%
\begin{array}{ccc}
 A&0 &B \\
 0&C&0  \\
 B &0 &A \\
\end{array}%
\right),
\eea
where
\bea
A&=&\fr 1 2 \left(a_1 \al_1+a_2 \al_2\right), \,\, B=\frac{1}{2}\left(a_1 \beta_1+a_2 \beta_2\right),\crn
C&=&\fr{a_2 b_2 g_1-a_2 b_1 c_1 - a_1 b_2 d_1}{c_1 d_1}-\fr{(a_1 d_1 - a_2 g_1) (b_1 c_1 - b_2 g_1)}{g_1^2 g_3},\label{ABC}\eea
\bea
\al_1&=& -\fr{2 b_2 c_1}{c_1^2 - c_2^2}+\fr{(d_1 - d_2) [b_2 (g_1 - g_2)-b_1 (c_1-c_2)]}{(g_1 - g_2)^2 (g_3 - g_4)}
+\fr{(d_1 + d_2) [b_2 (g_1 + g_2)-b_1 (c_1 + c_2)]}{(g_1 + g_2)^2 (g_3 + g_4)},\crn
\al_2&=&-\fr{2 b_1 d_1}{d_1^2 - d_2^2}-\fr{2 b_2 [(c_1 d_1 + c_2 d_2) g_1 - (c_2 d_1 + c_1 d_2) g_2]}{(c_1^2 - c_2^2) (d_1^2 - d_2^2)}+\fr{b_1 (c_1 - c_2)}{(g_1 - g_2) (g_3 - g_4)} \crn
&+&\fr{b_1 (c_1 + c_2)}{(g_1 + g_2) (g_3 + g_4)}-\fr{2 b_2 g_3}{g_3^2 - g_4^2}, \crn
\beta_1&=&\fr{2 b_2 c_2}{c_1^2 - c_2^2}+\fr{(d_1 - d_2) [b_1 (c_1 - c_2) - b_2 (g_1 - g_2)]}{(g_1 - g_2)^2 (g_3 - g_4)}+\fr{(d_1 + d_2) [b_2 (g_1 + g_2)-b_1 (c_1 + c_2)]}{(g_1 + g_2)^2 (g_3 + g_4)}, \crn
\beta_2&=&\fr{2 b_1 d_2}{d_1^2 - d_2^2}+\fr{2 b_2 [(c_1 d_1 + c_2 d_2) g_2-(c_2 d_1+ c_1 d_2) g_1 ]}{(c_1^2 - c_2^2) (d_1^2 - d_2^2)}
-\fr{b_1 (c_1 - c_2)}{(g_1 - g_2) (g_3 - g_4)}\crn
&+&\fr{b_1 (c_1 + c_2)}{(g_1 + g_2) (g_3 + g_4)}+\fr{2 b_2 g_4}{g_3^2 - g_4^2},
\eea
with $a_{1,2}, b_{1,2}, c_{1,2}$, $d_{1,2}$ and $g_{1,2,3,4}$ defined in Eqs. (\ref{submatrix1}) - (\ref{submatrix6}).
The mass matrix $m_\nu $ for light active neutrinos is diagonalized by the rotation matrix $U_{\nu }$,
\bea
U_\nu &=& \left(%
\begin{array}{ccc}
 \fr{1}{\sqrt{2}}&0 &-\fr{1}{\sqrt{2}}  \\
 0                          &1 & 0  \\
\fr{1}{\sqrt{2}}&0 &\fr{1}{\sqrt{2}}  \\
\end{array}%
\right),\label{Unu}
\eea
and the light active neutrino masses $m_{1,2,3}$ are given by
\bea
 m_1& =&A+B, \hs  m_2= C,\hs  m_3 = A-B. \label{m123}
 \eea
By combining Eqs. (\ref{Uclep}) and (\ref{Unu}) we find that the leptonic mixing matrix takes the form:
\bea
U^{lep}&=&U_L^\+  U_\nu = \left(%
\begin{array}{ccc}
\fr{1 +e^{-i \beta}}{\sqrt{6}}&\fr{ e^{-i \al}}{\sqrt{3}} & \fr{-1 +e^{-i \beta}}{\sqrt{6}}  \\
 \fr{1 +\om^2 e^{-i \beta}}{\sqrt{6}}&\fr{\om  e^{-i \al}}{\sqrt{3}}& \fr{-1 +\om^2 e^{-i \beta}}{\sqrt{6}}   \\
 \fr{1 +\om e^{-i \beta}}{\sqrt{6}} & \fr{ \om^2 e^{-i \al}}{\sqrt{3}} & \fr{-1 +\om e^{-i \beta}}{\sqrt{6}}   \\
\end{array}%
\right).\label{Ulepg}
\eea
We see that all the elements of the matrix $U^{lep}$ in Eq. (\ref{Ulepg}) depend only on two parameters $\al$ and $\beta$. From
experimental constraints on the elements of the lepton mixing matrix given
in Ref. \cite{Uconstraint}, we can find out the regions
of $\al$ and $\beta$ to establish experimental constraints for lepton
mixing matrix.
In the standard Particle Data Group (PDG) parametrization, the leptonic mixing
 matrix can be parameterized in three Euler's angles as follows:
 \bea s_{13}&=&\left|U_{13}\right|=\fr{\sqrt{1 - \cos{\beta}}}{\sqrt{3}},\label{s13N}\\
 t_{23}&=&\left|\fr{U_{23}}{U_{33}}\right|=\left|\fr{1 + 2 \cos{\beta}}{2 + \cos{\beta} -\sqrt{3}\sin{\beta}}\right|,\label{t23N}\\
  t_{12}&=&\left|\fr{U_{12}}{U_{11}}\right|=\sqrt{\fr{1}{1+\cos \beta}}, \label{t12N}
\eea
i.e, $s_{13}, t_{12}$ and $t_{23}$ in Eqs. (\ref{s13N}) and (\ref{t12N}) depend only on one parameter $\beta$.
Eqs. (\ref{s13N}) - (\ref{t12N}) yields:
\bea
\beta&=&- \mathrm{arccos}{(1 - 3 s_{13}^2)}, \label{betaN1} \\
t_{23}&=&\fr{1 - 2 s_{13}^2}{1 - s_{13}^2+s_{13}\sqrt{2 - 3 s_{13}^2}}, \label{t12N1}\crn
t_{12}&=&\fr{1}{\sqrt{2-3 s^2_{13}}}. \label{t23N1}
\eea
\begin{figure}[h]
\begin{center}
\includegraphics[width=8.0cm, height=5.5cm]{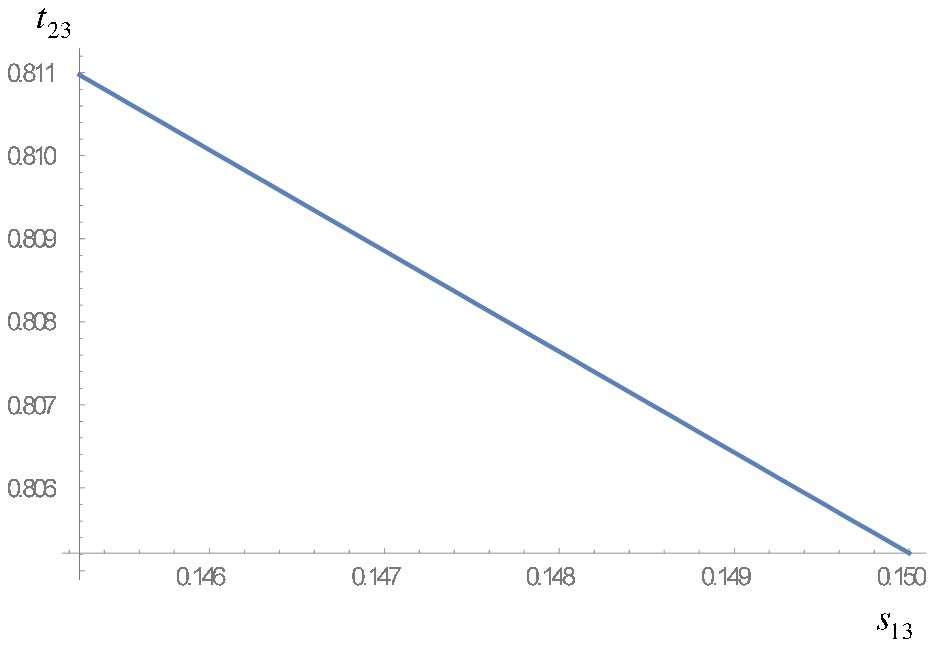}
\vspace*{-0.1cm} \caption[$t_{23}$ as a function of
 $s_{13}$ with $s_{13}\in (\sqrt{0.0211},\sqrt{0.0225})\, \mathrm{rad}$]{$t_{23}$ as a function of
 $s_{13}$ with $s_{13}\in (\sqrt{0.0211},\sqrt{0.0225})\, \mathrm{rad}$.}\label{t23}
\vspace*{-0.1cm}
\end{center}
\end{figure}
\begin{figure}[h]
\begin{center}
\includegraphics[width=8.0cm, height=5.5cm]{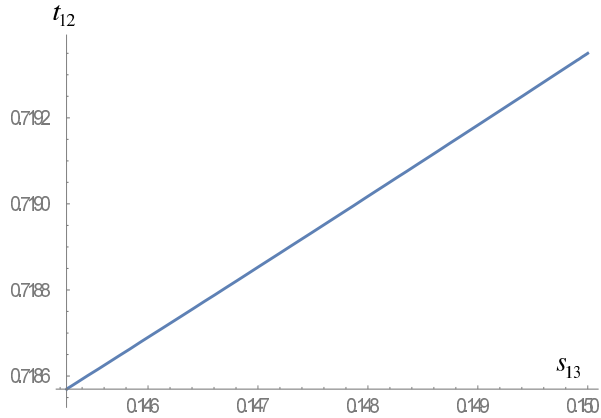}
\vspace*{-0.1cm} \caption[$t_{12}$ as a function of
 $s_{13}$ with $s_{13}\in (\sqrt{0.0211},\sqrt{0.0225}) \,\mathrm{rad}$]{$t_{12}$ as a function of
 $s_{13}$ with $s_{13}\in (\sqrt{0.0211},\sqrt{0.0225}) \,\mathrm{rad}$.}\label{t12}
\vspace*{-0.1cm}
\end{center}
\end{figure}
The data in Particle Data Group 2018 \cite{PDG2018} shows that $s_{13}\in (0.145258, 0.15) \,\mathrm{rad}$ so $t_{23}\in (0.806, 0.811)$ and $t_{12} \in (0.7811, 0.7192)\,\mathrm{rad}$ as depicted in Figs. \ref{t23} and \ref{t12}, respectively. Taking the best fit value given in Ref. \cite{PDG2018}, $s_{13}=0.147648\, \mathrm{rad} \, (\theta_{13}=8.45963^\circ)$ we get $t_{23}=0.808068$ ( $\theta_{23}=38.9406^\circ )$ and $t_{12}=0.718959$  ($\theta_{12}=35.7146^\circ)$ which are in good agreement with the values of $\theta_{23}$ and $\theta_{12}$ given in Ref. \cite{PDG2018}. On the other hand, with this best value of $\theta_{13}$, we get $\beta =-0.363663 \, \mathrm{rad} \, (\sim 339.163^\circ)$ and Dirac CP violation phase $\de _{CP}=259.579^\circ$ which is a viable value of the CP violating Dirac phase \cite{PDG2018}. The leptonic mixing matrix in Eq. (\ref{Ulep}) takes the explicit form
\bea
U^{lep}= \left(%
\begin{array}{ccc}
0.789797 + 0.145214 i   & 0.57735 e^{-i a}                  &-0.0266994 + 0.145214 i  \\
0.343233 - 0.403038 i   & (-0.288675 + 0.5 i) e^{-i a}  & -0.473264 - 0.403038 i  \\
0.0917147 + 0.257824 i & (-0.288675 - 0.5 i) e^{-i a}  & -0.724782 + 0.257824 i  \\
\end{array}%
\right),\label{Ulep}
\eea
which is an unitary matrix.

The expression (\ref{Ulep}) shows that $\al$ is free parameter so we can choose the VEV alignment $\phi$ in the charged-lepton sector as $\langle\phi \rangle= v(1, 1, e^{i\beta})$, i.e, $\al$ may get the value $\al=0$. In this case, the leptonic mixing matrix becomes:
\bea
U^{lep}= \left(%
\begin{array}{ccc}
0.789797 + 0.145214 i   & 0.57735                &-0.0266994 + 0.145214 i  \\
0.343233 - 0.403038 i   & -0.288675 + 0.5 i  & -0.473264 - 0.403038 i  \\
0.0917147 + 0.257824 i & -0.288675 - 0.5 i  & -0.724782 + 0.257824 i  \\
\end{array}%
\right),\label{Ulep0}
\eea
or
\bea
|U^{lep}|&=& \left(%
\begin{array}{ccc}
0.803036 & 0.57735 & 0.147648 \\
0.529385 &0.57735 & 0.621625 \\
0.273651 &0.57735 & 0.769274 \\
\end{array}%
\right),\label{Ulep1}
\eea
i.e, the ranges of the magnitude of the elements of the three-flavour
leptonic mixing matrix  is consistent with those of given in Ref. \cite{Uconstraint}.
At present, the values of neutrino masses (or the absolute neutrino
masses) as well as the mass ordering of neutrinos are still unknown. The result in Ref. \cite{Tegmark} shows that $m_i\leq 0.6\, \mathrm{eV} \, ( i=1,2,3)$ while the upper bound on the sum of light active neutrino masses is given by \cite{constraint},
\bea
\sum^3_{i=1} m_i \leq  0.17 \, \mathrm{eV}. \label{sum}
\eea
 The experimental neutrino oscillation data given in Eq. (\ref{PDG2018}) are compatible with two possible signs of
$\De  m_{23}^{2}$ which is currently unknown and correspond to two types of neutrino mass spectra.
\subsection{Normal spectrum ($m_{1}< m_{2}<m_{3}$)}
By taking the best fit values on neutrino mass squared differences for normal spectrum, given in Ref. \cite{PDG2018},
 $\De  m_{21}^2 =7.53\times 10^{-5} \mathrm{eV}^2$ and $\De  m_{32}^2 =2.444\times 10^{-3}\mathrm{eV}^2$, we obtain four solutions, however, they have the same absolute values of $m_{1,2, 3}$, the unique
difference is the sign of them. So, here we only consider the following solution:
\bea
A &=& 1.58114\times 10^{-2} \Ga , \crn
B &=&  \left(-0.0148662-12.5522 C^2+7.93871\times 10^{-3} \ga\right) \Ga ,\label{AB}
\eea
where
\bea
\Ga  &=& \sqrt{2.3687+2\times 10^3 C^2+1.26491\sqrt{\ga  }},\crn
 \ga  &=& \sqrt{-0.460083 +5.92175\times 10^3 C^2 + 2.5\times 10^6 C^4} . \label{gama}
\eea
In the model under consideration, $C\equiv m_2 \in (0.001, 0.0506)\, \mathrm{eV}$ is a good region of $C$ that can reach the realistic normal neutrino mass
hierarchy which is depicted in Fig. \ref{m13}. In the case $C\equiv m_2=0.0087\, \mathrm{eV}$, the parameters $A,B$ and the other neutrino masses are explicitly
 given as $A=2.54105\times 10^{-2}, B=-2.4786\times 10^{-2}$, $m_1=6.245\times 10^{-4}\, \mathrm{eV}$ and $m_3=5.01965\times 10^{-2}\, \mathrm{eV}$ which corresponds to a normal
neutrino mass spectrum. The sum of all three neutrino in this case is given by $\sum^N=\sum_{i=1}^3 m_i = 5.9521\times 10^{-2} \mathrm{eV}$ lying within the cosmological
bound from Planck data given in Eq. (\ref{sum}).
\begin{figure}[h]
\begin{center}
\includegraphics[width=7.5cm, height=5.5cm]{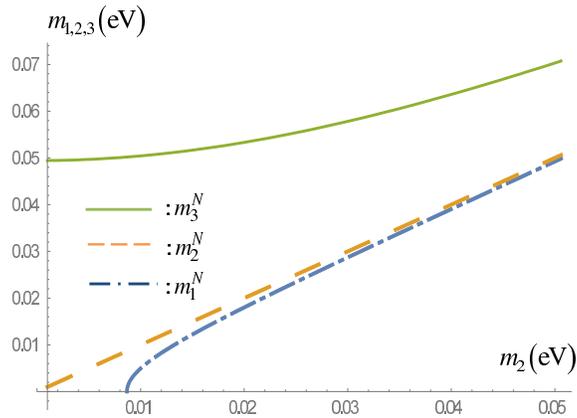}
\vspace*{-0.1cm} \caption[$m_{1,3}$ as functions of
 $m_2 $ with $m_2 \in (0.001, 0.0506) $ in the normal spectrum]{$m_{1,3}$ as functions of
 $m_2 $ with $m_2 \in (0.001, 0.0506)$ in the normal spectrum.}\label{m13}
\vspace*{-0.1cm}
\end{center}
\end{figure}
\begin{figure}[h]
\begin{center}
\includegraphics[width=7.5 cm, height=5.5cm]{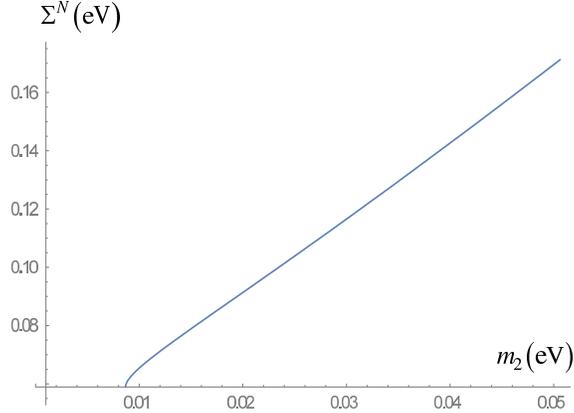}
\vspace*{-0.1cm} \caption[$\sum^N=\sum_{i=1}^3 m^N_i$ as functions of
 $m_2 $ with $m_2 \in (0.001, 0.0506) $ in the normal spectrum]{$\sum^N=\sum_{i=1}^3 m^N_i$ as functions of
 $m_2 $ with $m_2 \in (0.001, 0.0506)$ in the normal spectrum.}\label{m13}
\vspace*{-0.1cm}
\end{center}
\end{figure}
\subsection{Inverted spectrum ($m_{3}< m_{1}<m_{2}$)}
\begin{figure}[h]
\begin{center}
\includegraphics[width=7.5cm, height=5.5cm]{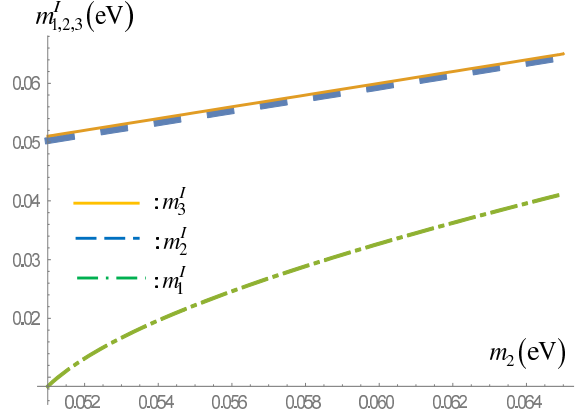}
\vspace*{-0.1cm} \caption[$m^I_{1,3}$ as functions of
 $C\equiv m_2 $ with $m_2 \in (0.051, 0.065) $ in the inverted spectrum]{$m^I_{1,3}$ as functions of
 $C\equiv m_2 $ with $m_2 \in (0.051, 0.065) $ in the inverted spectrum.}\label{m123I}
\vspace*{-0.1cm}
\end{center}
\end{figure}

Similar to the normal spectrum, by taking the best fit values on neutrino mass squared differences for inverted spectrum, given in Ref. \cite{PDG2018},
 $\De  m_{21}^2 =7.53\times 10^{-5} \mathrm{eV}^2$ and $\De  m_{32}^2 =-2.53\times 10^{-3}\mathrm{eV}^2$, we get a solution as follows:
\bea
A &=& 1.5811\times 10^{-2} \Ga' , \crn
B &=&  \left(-0.0167814+12.8825 C^2-1.2882\times 10^{-2} \ga'\right) \Ga' ,\label{AB}
\eea
where
\bea
\Ga' &=& \sqrt{2.6053+2\times 10^3 C^2+2\sqrt{\ga'}},\crn
 \ga' &=& \sqrt{0.190509-2.6053\times 10^3 C^2 + \times 10^6 C^4} . \label{gama}
\eea
In this model, $C\equiv m_2 \in (0.051, 0.065)\, \mathrm{eV}$ is a good region of $C$ that can reach the inverted neutrino mass hierarchy which is depicted in Fig. \ref{m123I}. In the case $C\equiv m_2=5.1\times 10^{-2} \, \mathrm{eV}$, the parameters $A,B$ and the other neutrino masses are explicitly
 given as $A=2.93412\times 10^{-2}, B=2.09151\times 10^{-2}$, $m^I_1=5.02563\times 10^{-2}\, \mathrm{eV}$ and $ m^I_3=8.42615\times 10^{-3}\, \mathrm{eV}$ which corresponds to an inverted
neutrino mass spectrum. The sum of all three neutrino in this case is given by $\sum^I=\sum_{i=1}^3 m^I_i = 0.10968\, \mathrm{eV}$ which is consistent with the cosmological
bound from Planck data in Eq.\eqref{sum}.
\begin{figure}[h]
\begin{center}
\includegraphics[width=7.5 cm, height=5.5cm]{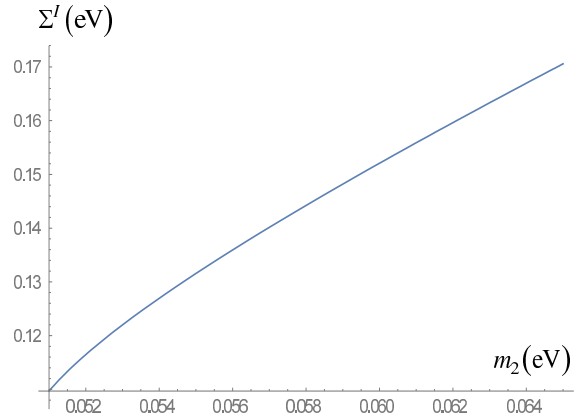}
\vspace*{-0.1cm} \caption[$\sum^I=\sum_{i=1}^3 m^I_i$ as functions of
 $m_2 $ with $m_2 \in (0.051, 0.065) $ in the inverted spectrum]{$\sum^I=\sum_{i=1}^3 m^I_i$ as functions of
 $m_2 $ with $m_2 \in (0.051, 0.065) $ in the inverted spectrum.}\label{m1m2m3sI}
\vspace*{-0.1cm}
\end{center}
\end{figure}
\section{\label{conclusion}Conclusions}
We have proposed a low-scale seesaw model to generate the masses for the active neutrinos based on $S_4$ flavour symmetry supplemented by the $Z_2 \times Z_3 \times Z_4 \times Z_{14}\times U(1)_L$ group,
where the masses of the SM charged fermions and the fermionic mixing angles are generated from a Froggatt-Nielsen mechanism after the spontaneous breaking of the $S_4\times Z_2 \times Z_3 \times Z_4 \times Z_{14}\times U(1)_L$ group.
The obtained values for the physical observables of the quark and
  lepton sectors are in good agreement with the most recent experimental data. The Dirac CP violating phase $\de _{CP}$ is predicted to be $259.579^\circ$ which is consistent with the most recent neutrino oscillation experimental data \cite{PDG2018}. The predictions for the absolute neutrino masses in the model can also saturate the recent constraints.

\section*{Acknowledgments}
This research is funded by Vietnam National Foundation for Science and Technology Development (NAFOSTED) under grant number 103.01-2017.341 as well as by Fondecyt (Chile), Grants
No.~1170803, CONICYT PIA/Basal FB0821. H. N. L acknowledges the warm hospitality at BLTP, JINR and  the financial support of the
Vietnam Academy of Science and Technology under grant  NVCC05.01/19-19. A.E.C.H is very grateful to the Institute of Physics, Vietnam Academy
of Science and Technology for the warm hospitality.
\appendix

\end{document}